\documentclass[aps,prb,groupedaddress,twocolumn,10pt]{revtex4-1}

\begin{document}

\newcommand{\be}{\begin{equation}}
\newcommand{\ee}{\end{equation}}
\newcommand{\bea}{\begin{eqnarray}}
\newcommand{\eea}{\end{eqnarray}}

\title{Taking a critical look at holographic critical matter}

\author{D. V. Khveshchenko}

\affiliation{Department of Physics and Astronomy, University of North Carolina, Chapel Hill, NC 27599}

\begin{abstract}
Despite a recent flurry of applications of the broadly defined ('non-AdS/non-CFT') holographic correspondence to a variety of condensed matter problems, the status of this intriguing, yet  speculative, approach remains largely undetermined. This note exposes a number of potential inconsistencies between the previously made holographic predictions and advocates for a compelling need to systematically contrast the latter against the results of alternate, more conventional, approaches as well as experimental data. It is also proposed to extend the list of computed observables and utilize the general relations between them as a further means of bringing the formal holographic approach into a closer contact with the physical realm.
\end{abstract}

\maketitle

{\it Introduction}

Quantum many-body theory has long been 
seeking to expand its toolbox of computational techniques, thus allowing one to describe and classify a broad variety of non-Fermi liquid (NFL) states of strongly correlated fermions. 

In generic $1d$ fermion systems, the conventional Fermi liquid behavior gets (marginally) destroyed by an arbitrarily weak short-ranged repulsive interaction, thereby giving way to the so-called Luttinger behavior. As one of the hallmarks of the Luttinger regime, the electron propagator exhibits an algebraic decay with distance/time, $G(x)\propto 1/x^{\Delta}$, governed by an anomalous dimension $\Delta>1$. 
Moreover, long-range interactions, such as Coulomb, modify the Fermi liquid behavior even more drastically, resulting in, e.g., the $1d$ Wigner crystal state where the fermion propagator 
decays faster than any power law, $G(x)\sim\exp(-{\#}\ln^{3/2}x)$.

In higher dimensions, the Fermi liquid is generally believed to be more robust, although it is not expected to remain absolutely stable. While in the case of short-ranged repulsive interactions any departures from the Fermi liquid are likely to be limited to the strong-coupling regime, long-ranged couplings can possibly result in the NFL types of behavior without any threshold.

Of a particular interest are the spectroscopic and transport properties 
of such emergent critical behaviors as incipient $s$-, $p$-, and $d$-wave charge/spin density waves and orbital current-type instabilities in itinerant (anti)ferromagnets, quantum spin liquids, compressible ('composite fermion') Quantum Hall states, etc. 
Recently, the focus has also been on the $d>1$-dimensional zero density ('neutral') 
Dirac/Weyl systems characterized by the presence of isolated points ('nodes') or lines ('arcs') of vanishing quasiparticle energy. 

The intrinsic complexity of these systems has long been recognized, prompting the use of such sophisticated techniques as renormalization group, $1/N$- and $\epsilon$-expansions, Keldysh functional integral and quantum kinetic equation, supersymmetric 
diffusive and ballistic $\sigma$-models, multi-dimensional bosonization, etc.
In spite of all the effort, however, the overall progress towards a systematic classification of various 'strange' metallic (compressible) states that are often indiscriminately referred to as 'higher dimensional Luttinger liquids' has been rather slow. 

In that regard, the recent idea of a 
(broadly defined) holographic correspondence \cite{AdS}
could provide a sought-after powerful alternative technique. Specifically, its widely used 'bottom-up' version could potentially offer an advanced phenomenological framework for discovering new and classifying the already known types of NFL behavior. 

Although in much of the pertinent 
literature the validity of the generalized ('non-AdS/non-CFT') holographic conjecture appears to be taken for granted, it might be worth reminding that the actual status of the entire holographic approach remains anything but firmly established.

Indeed, in most of its applications this bold adaptation of the original 'bona fide' string-holographic correspondence does not seem to be subject to much (or, for that matter, any) of the former's stringent symmetry conditions, as the pertinent non-relativistic systems  at finite density and temperature, in general, tend to be neither non-Abelian/multicomponent/supersymmetric, nor even Lorentz invariant. 

The precious few examples of a quantitative agreement between the holographic approach and other (e.g., Monte Carlo) techniques involve some carefully 
tailored gravity duals (whose physical nature still remains rather obscure, though)  \cite{1309.2941,1309.5635}. 

In other cases, under a closer inspection the purported agreement 
appears to be largely limited to an apparent similarity between the results 
of some (for the most part, numerical) calculations and certain selected 
sets of the available experimental data. 

For one, in Ref.\cite{1204.0519} the holographically 
computed optical conductivity was claimed to agree 
(over less than half of a decade, $2<\omega\tau<8$) with
the enigmatic power-law decay, $\sigma(\omega)\sim\omega^{-2/3}$, 
observed in the normal state of the superconducting cuprates
($BSCYCO$), pnictides,
and certain heavy fermion materials, often up to the energies of
order $eV$. Notably, the original claim was not corroborated by the later 
analysis of Ref.\cite{1311.3292} and was also argued to be intermittent with the 'more universal' $\sim 1/\omega$ behavior \cite{1208.4102}.

Also, while being customarily wordy and profuse on technical details,
most of the works on holography end up with rather simple scaling
relations as their final answers, thereby suggesting that there might be more 
economic and physically illuminating ways of obtaining such results.  

Thus, in order for its status to be definitively ascertained, the holographic approach needs to be assessed critically and applied to those systems where a preliminary insight can be (or has already been) gained by some alternative means, so that a systematic comparison with the holographic results can be made. Also, in order to gain a predictive power the holographic calculations would have to be made for as many observables as possible and then applied to the host of experimental data on the documented NFL materials. 

The present communication takes another step towards filling in this gap.

{\it Practical holography of condensed matter systems}

In its original formulation, the holographic principle postulates that certain 
$d+1$-dimensional 'boundary' field theories allow for  
a dual description involving, alongside other 'bulk' fields, $d+2$-dimensional gravity. Moreover, when the boundary theory is strongly coupled, the higher-dimensional gravity appears to allow for a semiclassical treatment, thus facilitating a powerful new approach to the problem of strong interactions.

So far, the holographic approach has been opportunistically 
applied to a variety of systems which includes 'strange' Fermi and Bose metals describing quantum-critical  $U(1)$ and $Z_2$ spin liquids, itinerant (anti)ferromagnets, quantum nematics, Mott transitions in lattice and cold atom systems, Hall effect, graphene, etc. 
  
On the gravity side, the system in question  
would be characterized by a (weakly) fluctuating background metric
$g_{\mu\nu}=g^{(0)}_{\mu\nu}(r)+\delta g_{\mu\nu}(t,{\vec x},r)$
determining the interval 
\begin{equation}
ds^2=g_{tt}dt^2+g_{rr}{dr^2}+\sum_{ij}g_{ij}dx^idx^j
\end{equation} 

The early applications of the holographic approach  
revolved around a handful of the classic
'black brane' solutions, such as the Reissner-Nordstrom AdS (anti-de-Sitter) black hole
with the metric 
\be
g_{tt}\sim -{f(r)\over r^2},~~~ g_{rr}\sim {1\over r^2f(r)},~~~
g_{ij}(r)\sim\delta_{ij}{1\over r^2}
\ee
where the emblackening factor $f(r)$ 
vanishes at the horizon of radius $r_h$ which is inversely proportional
(ostensibly, similar to the case of the Schwarzschild black hole in the 
asymptotically Minkowski space-time, despite the variable's $r$ being 
the $inverse$ of the actual radius in the $d+1$-dimensional bulk space) 
to the Hawking temperature $T$ 
shared by the bulk and boundary degrees of freedom. The explicit 
form of this function depends on how the gravito-electro-magnetic 
background is described. 

In the black brane geometry of the minimal
Einstein-Maxwell theory, one has $f_{EM}(r)=1-(r/r_h)^{d+1}$, 
whereas in the Dirac-Born-Infeld (DBI) theory with the Lagrangian 
\be
L_{DBI}={\sqrt {-det[g_{\mu\nu}+F_{\mu\nu}]}}
\ee
geared to the strong-field limit $f_{DBI}(r)={\sqrt {1+(en)^2r^{2d}}}$, $en$ being the 
boundary density of electric charge.

In the early works on the subject, the metric (2) was claimed to provide a potential  
gravity dual to the class of strongly correlated condensed matter problems - most notably, heavy fermion materials and cuprates - which are believed to manifest a certain 'semi-locally critical' behavior \cite{semilocal}. However, soon thereafter it was realized that the corresponding  physical scenario appears to be much too limited to encompass more general types of 
the real-life NFLs, so the focus shifted towards a broader class of geometries.

Further attempts of 'reverse engineering' have brought out such Lorentz non-invariant metrics as the Shroedinger, Lifshitz, helical Bianchi, etc.
Amongst them, a particular attention has been paid to the 
static, diagonal, and isotropic metrics with algebraic radial dependence 
\be 
g_{tt}\sim -{1\over r^{2\alpha}},~~~g_{rr}\sim {1\over r^{2\beta}}, ~~~ g_{ii}\sim {1\over r^{2\gamma}}
\ee 
These exponents are defined modulo a change of the radial 
variable $r\to\rho^\delta$ resulting in the substitution 
\be 
\alpha\to\alpha\delta,~~~ \beta\to\beta\delta-\delta+1,~~~
\gamma\to\gamma\delta
\ee
Unless $\gamma=0$, the metric (4) is conformally  
equivalent to the one 
\be 
g_{tt}\sim -r^{2(\theta/d-z)},~~~g_{rr}=g_{ii}\sim r^{2(\theta/d-1)}
\ee
characterized by only two parameters
\be
\theta=d{1-\beta\over 1-\beta+\gamma},~~~
z={1+\alpha-\beta\over 1-\beta+\gamma}
\ee

which describe a family of  
'hyperscaling-violating' (HV) backgrounds \cite{hyperscaling} 
where the dynamical exponent $z$ controls the boundary excitation 
spectrum $\omega\propto q^z$, while $\theta$ quantifies  
a non-trivial scaling of the interval $ds\to \lambda^{\theta/d} ds$,
the scaling-(albeit not Lorentz-) invariant ('Lifshitz') case corresponding to $\theta=0$.

The finite-$T$ version of the HV metric can be constructed 
by decorating (6) with the additional
factor $f_{HV}(r)=1-(r/r_h)^{d+z-\theta}$, akin to Eq.(2),
which introduces the black brane's horizon located at $r_h\sim T^{-1/z}$.

The physically sensible values of $z$ and $\theta$ are expected to satisfy
the all-important 'null energy conditions'
\be
(d-\theta)(d(z-1)-\theta)\geq 0,~~~(z-1)(d+z-\theta)\geq 0
\ee
signifying a thermodynamic stability of the corresponding geometry.

The HV metrics have been extensively discussed in the content of various generalized gravity theories, including those with massive vector fields
as well as the Einstein-Maxwell-dilaton (EMD) 
theory which includes an additional scalar field, alongside the cosmological constant term  \cite{hyperscaling} 
\begin{equation}
L_{EMD}={1\over 2\kappa^2}(R+{d(d+1)\over L^2})-{(\partial\phi)^2\over 2}-U(\phi)
-{V(\phi)\over 2e^2}F_{\mu\nu}^2
\end{equation}
In its minimal version, both, the dilaton potential $U(\phi)$ and the effective gauge coupling $V(\phi)$ are given by some exponential functions of $\phi$. 

At the (semi)classical level, gravitating matter 
added to the EMD Lagrangian (9) 
can be described in terms of its energy-momentum tensor
and electric current
$$
T_{\mu\nu}=(E+P)u_{\mu}u_{\nu}+Pg_{\mu\nu},~~~J_{\mu}=enu_{\mu}
$$
where $en, E, P, u_{\mu}$ are the charge and energy densities, pressure, 
and (covariant) local velocity, respectively. 
The $1 st$ Law of thermodynamics then relates the above quantities as follows  
\be 
E+P=ST+\mu n
\ee
where $S$ is entropy density and $\mu=eA_t(r)|_{r\to 0}$ is the chemical potential. In the particle-hole symmetric ('neutral') system $\mu=0$ and the equation of state reads $E=(d-\theta)P/z$. 

The HV solutions (6) have also been obtained by 
taking into account a back-reaction of the matter on the background geometry \cite{electronstar}.
Such analyses would typically use the Fermi distribution 
when summing over the occupied 
fermion states, thereby achieving a partial account of  
the (Hartree-type) effects of the Fermi statistics, 
while leaving out more subtle (exchange and correlation) ones.

However, for an already chosen gravitational background the customary way of introducing a finite charge density into the holographic scheme is by embedding a D-brane into such geometry and treating it in the probe approximation (no back-reaction). The pertinent dynamics is 
then described by the DBI action (3) with the background electric field  
\be 
F_{rt}=\partial_rA_t=
{\sqrt {|g_{tt}|g_{rr}(en)^2\over \prod_ig_{ii}+(en)^2}}
\ee

The DBI approach has been used to study thermodynamics of the HV theory.
In that regard, in Refs.\cite{1306.3816} the specific heat  
of a finite density ('charged') system was found to scale with temperature as 
\be
C_{DBI,charged}\sim T^{-{2\theta/zd}}
\ee
Being primarily interested in the limit $z\to\infty, -\theta/z\to const$, 
the authors did not seem to be particularly concerned with 
the implications of this result, including its apparent inapplicability in 
the potentially physically relevant case of $\theta=d-1>0$ (see below).

Moreover, even for  $\theta=0$ Eq.(12) differs from the 
expression obtained in the earlier work 
of Refs.\cite{1007.0590} where the standard ('black-body') 
leading term $\sim T^{d/z}$ was deliberately discarded 
in favor of the subdominant (yet, charge density dependent) one, $\sim T^{2d/z}/en$.

Nevertheless, Eq.(12) can be rationalized by
comparing it to the result of a direct calculation for the HV metric (6)
and $\mu=0$   
\bea
C_{DBI,neutral}\sim {\partial\over \partial T}\int^{r_h}_0 dr {\sqrt {-det g_{\mu\nu}}}\nonumber\\
\sim T^{(d-\theta-2\theta/d)/z}
\eea
Physically, this expression can also be recognized 
as the ($number$, rather than $charge$) density $n(T)$ of thermally excited carriers
(of either sign). 

One subtle point is that in the charged case
it is not the latter but the $charge$ density that gets replaced with 
a finite value at not too high temperatures.
Based on that insight, the entire
temperature dependence in Eq.(12) should then be attributed 
to the effective ($T$-dependent) charge 
\be
e\sim T^{2\theta/dz}
\ee
while the $charge$ density itself 
scales as the ratio between Eqs.(12) and (13) 
\be
en=C_{DBI,neutral}/C_{DBI,charged}\sim T^{(d-\theta)/z}
\ee
The temperature dependence (13) would also be shared by 
the concomitant thermal entropy 
\be
S_{DBI}\sim n\sim T^{(d-\theta-2\theta/d)/z}
\ee
Notably, this result of a straightforward thermodynamic calculation
appears to be at odds with both, the naive estimate 
$$
S_{BH}={1\over 4}A\sim r_h^{-d}\sim T^{d/z}
$$
for the Bekenstein-Hawking entropy a black $d$-brane of radius $r_h$
as well as the entanglement entropy 
\be
S_{ent}\sim T^{(d-\theta)/z}
\ee
which relation would often be quoted $ad~hoc$ with regard to the empiric interpretation  
of the parameter $\theta$ as a 'dimensional defect' that gives rise to the effective dimension 
$d_{eff}=d-\theta$ in the above expression for $S_{BH}$.

The proper choice of $\theta$ has been extensively discussed 
in the context of fermionic entanglement entropy 
which points to the value $\theta=d-1$, consistent with the notion of the Fermi surface as a $d-1$-dimensional manifold spanning the tangential directions in the reciprocal (momentum) space of the boundary theory \cite{1308.3234}. 
 
In fact, by adhering to the above value one chooses to treat the HV system in question as fermionic and, therefore, must use the fermion quasiparticle dispersion  
$\omega\sim(k-k_F)^{z_f}$ to determine the value of the dynamical index.
In what follows, the metric with the parameters $z=z_f,~\theta=d-1$
will be called 'Model I'.

Alternatively, one could treat the system as bosonic and use the value 
$z_b$ deduced from the dispersion of the bosonic mode, $\omega\sim k^{z_b}$.
A straightforward choice for the HV parameter would then be $\theta_b=0$, thereby reducing the corresponding metric back to the Lifshitz one (hereafter, 'Model II').

Yet another possibility corresponds to choosing $z=z_f$
and $\theta=d(1-z_f/z_b)$, which choice will be called 'Model III'. 
In fact, the two latter metrics are related by a conformal 
transformation, so some of the results turn out to be the same in both cases.
And, lastly, there also exists a choice $z=z_b$
and $\theta=d-z_b/z_f$ which will be referred to as 'Model IV'.

{\it Scaling properties of hyperscaling-violating systems}

The above exposition of the physically incomprehensible 
holographically computed specific heat suggests 
that any substantive comparison with 
the calculations performed by alternate techniques 
has to involve more than one quantity. 

Indeed, a power-law behavior of the specific heat is just one of 
the many scaling laws which describe quantum-critical systems. 
The list of other observables includes (tunneling) density of states, charge, current, and spin susceptibilities, electrical, thermal, and spin conductivities, shear and bulk viscosities, etc. 

In the quantum-critical regime of a massless ($m=0$)
and particle-hole symmetric (or neutral, $\mu=0$) 
system, the single most important scale is set by temperature $T$
or frequency $\omega$, whichever is greater.

The anticipated algebraic behavior of a physical 
observable $A$ is then fully characterized 
by its scaling dimension $[A]$, namely: 
$$
A(\omega, T)\sim max|T,\omega|^{\Delta_A},~~~~~\Delta_A=[A]/z
$$
Although in the following discussion no distinction is made between the exponents controlling the frequency and temperature dependencies, this point will be addressed later.

Once a new scale enters the game, the pure algebraic dependencies would only hold at high enough frequencies and/or temperatures, while at smaller $\omega$ or $T$ any pure power-law gets complemented by a universal function 
of the ratios between $T$ and all the competing scales ($m, \mu$, etc.). 

Also, in the case of a vanishing exponent 
one can encounter a logarithmic behavior $\sim\log max|T,\omega|$
stemming from, e.g., quantum localization or the well-known classical 'long-tail' phenomenon \cite{longtail}.

The scaling analysis begins with a proper assignment of the 
scaling dimensions under transformation of the space-time 
coordinates in the boundary theory.  

In accordance with the underlying dispersion relation ($\omega\sim k^z$),
the dimensions of the space-time coordinates and their conjugate 
energies/momenta take the values 
\be
[x_i]=-[k_i]=-1,~~[t]=-[\epsilon]=-[\mu]=-[T]=-z
\ee
whereas those of the gauge potential differ from the above values  
by the dimension of the effective charge (14)
\be
[A_i]=[k_i/e]=1-2\theta/d,~~~[A_0]=[\mu/e]=z-2\theta/d
\ee

From that one can also obtain the scaling relations 
$$
[v_i]=[x_i/t]=z-1,~~[\nabla_iT]=z+1
$$
$$
[{\cal E}_i]=[A_i/t]=[A_0/x_i]=z+1-2\theta/d
$$
\be
[B_i]=[A_j/x_k]=2-2\theta/d
\ee
where $v_i, A_i, {\cal E}_i, B_i$ are the velocity, vector potential, 
electric and magnetic fields, respectively. 

The dimensions of the 
energy and number densities can be read off directly from Eq.(13)
\be
[E]=[P]=[T_{tt}]=[T_{ij}]=[\mu n]=z+d-\theta-2\theta/d
\ee
Then, in order for the boundary action to maintain scale invariance,
a spatial integration must be thought of as contributing the extra dimension 
\be
[{\bf dx}]=-[n]=-d+\theta+2\theta/d,  
\ee
under which convention the total (quasi)particle number
$\int n{\bf dx}$ is dimensionless and, therefore, conserved.
In contrast, the total charge $\int en {\bf dx}$ appears to scale with the non-vanishing dimension (14) imposed by the 'running' dilaton-dependent gauge coupling. 
For a more systematic derivation of such assignment the
scaling properties of the entire bulk theory (9) have to be considered \cite{dvk_new}. 

Likewise, the dimensions of the electrical current and the remaining components of the stress-energy tensor can be readily deduced from the conservation laws
$$
e{\partial n\over \partial t}+{\partial J_i\over \partial x_i}=0,~~
{\partial T_{tt}\over \partial t}+{\partial T_{it}\over \partial x_i}=0,~~
{\partial T_{ti}\over \partial t}+{\partial T_{ji}\over \partial x_j}=0
$$
In this way, one obtains 
$$
[J_i]=d+z-1-\theta,
$$
$$
[T_{ti}]=d-\theta+1-2\theta/d,
$$
\be
[Q_i]=[T_{it}]=d+2z-1-\theta-2\theta/d
\ee
where $Q_i$ is the thermal current.

It is also worth pointing out that in the absence of the Lorentzian symmetry 
the stress-energy tensor becomes non-symmetrical (cf. with Refs.\cite{1304.7481}
which dealt exclusively with the case of $\theta=0$, though).

The density susceptibility $\chi$ (related to the charge one by the factor $e^2$), 
electrical $\sigma$ and thermal $\kappa$ conductivities
are then characterized by the following dimensions
\be
[\chi]=[E/\mu^2]=d-z-\theta-2\theta/d,
\ee
\be
[\sigma]=[J_i/E_i]=d-2-\theta+2\theta/d,
\ee
\be
[\kappa]=[Q_i/\nabla_iT]=d+z-2-\theta-2\theta/d
\ee
Curiously enough, prior to the release of the original version of this note
such simple scaling relations do not seem to have appeared in the holographic analyses
for general values of $z$ and $\theta$.

In the recent Ref.\cite{Karch}, an attempt was made to
further generalize these scaling relations to the situation where 
the spatial dimension of the gauge degrees of freedom $d_s$ can 
be different from that of the gravitational ones.
 
In this approach, instead of attributing the 'subleading' term $\phi=2\theta/d$ 
appearing in the above expressions (cf. Eq.(3.6) in Ref.\cite{Karch}) 
to the dimension of the electric charge (14), thus distinguishing between 
the number and charge densities (or, for that matter, $\mu$ and $A_0=\mu/e$
which Ref.\cite{Karch} makes no distinction between), 
the gauge sector was assigned its own HV parameter $\theta_m$. 

Moreover, it was argued in Ref.\cite{Karch} that
introducing an extra parameter (either $\theta_m$ or $\phi$) 
in addition to $z$ and $\theta$  is necessary 
for the proper description of a charged system with the HV geometry. 
However, even after such a modification 
the approach of Ref.\cite{Karch} still struggles 
to reproduce the thermodynamics of the DBI system self-consistently. 

Specifically, while it asserts that the dimensions of the energy 
densities in the gravitational and gauge sectors can  
generally be different, it still seeks to make the two 
equal, provided that the following condition is satisfied 
\be
\theta=d-d_s+\theta_m+\phi
\ee 
(cf. Eq.(3.16) in Ref.\cite{Karch} where the implicit
assumption $d_s=d$ seems to have been made).

In fact, for $d=d_s$ and the original DBI action (3) this condition would be impossible to meet for any $\theta\neq 0$, given that the pertinent value of the gauge HV parameter is $\theta_{m,DBI}=d_s\theta/d$ (see Eq.(3.8) in Ref.\cite{Karch}). 

Among other things, such inference implies that the dimension of the gauge sector's entropy 
is $[S]=[E/T]=d_s-\theta_m-\phi$ (see Eq.(3.11) in Ref.\cite{Karch}) which agrees with
Eq.(16) for $d_s=d$, yet differs from the much-anticipated empirical dependence (17)
(in Ref.\cite{Karch}, the latter was instead postulated for the entropy of the gravitational sector). In any case, though, this extended scheme does not provide 
a suitable framework for assessing the status of the earlier DBI studies 
pertaining to the original (i.e., $2$-parameter) HV systems. 

In contrast, the discussion presented in this Section (specifically, for the case of $d_s=d$) 
involves just one type of entropy (16) 
and requires no additional parameters or some other sleight of hand. 

It is, therefore, quite remarkable
that despite such discrepancies both variants of the dimensional 
assignments result in the same Eq.(25) for the conductivity, thereby
attesting to the intrinsic robustness of this and similar expressions
(moreover, in the original version of this note the very same result  
was obtained in still another couple of different ways). 

Furthermore, even robuster than the individual thermodynamic and kinetic coefficients
are their universal ratios which are dimensionless and, therefore, scale invariant 
(hence, constant for $\mu\ll T$).  
Amongst those are the standard Wilson and Wiedemann-Franz ratios 
whose vanishing dimensions follow from the above Eqs.(13,14) and (24-26)
\be
{\chi T\over C}=const,~~~{e^2\kappa\over \sigma T}=const
\ee
Should, however, a new scale emerge, these ratios, albeit remaining dimensionless, would no longer remain constant. In fact, they may 
deviate strongly from their Fermi liquid 
values, thereby signalling, e.g., the formation of a strongly correlated (hydrodynamic) quantum-critical state.

In that regard, albeit being irrelevant for the general scaling properties, a practically important distinction has to be made between the formally defined thermal conductivity (see Eq.(30) below) and that computed under the condition of vanishing electric current (which setup more faithfully represents the actual measurement). 

As a means of lending further support to the above scaling relations, one can 
also reproduce the dimensions (25) and (26) of the 
electrical and thermal conductivities from the Kubo formulae
\be
\sigma={1\over \omega}Im\int {\bf dx}\int^{\infty}_{0}dt
e^{i\omega t}<[J(t,x),J(0,0)]>
\ee
and
\be
\kappa={1\over \omega T}Im\int {\bf dx}\int^{\infty}_{0}
dte^{i\omega t}<[Q(t,x),Q(0,0)]>
\ee
where the use the scaling rule (22) is instrumental. 

As yet another independent check, the shear viscosity 
$$
\eta={1\over \omega}Im\int {\bf dx}\int^{\infty}_{0}dte^{i\omega t}<[T_{xy}(t,x),T_{xy}(0,0)]>
$$
features the dimension 
$$
[\eta]=d-\theta-2\theta/d, 
$$ 
which, together with Eq.(16), guarantees 
that the celebrated viscosity-to-entropy ratio $\eta/S$ is indeed dimensionless.

It can also be easily seen that 
the dimensions are consistent with the classical (Einstein's) relations
\be
\sigma=e^2\chi D,~~~~~
\eta=D(E+P)/v^2
\ee
where the diffusion coefficient $D=v^2/d\Gamma$ contains a  
scattering rate $\Gamma$ expected to assume the universal linear form 
\be
\Gamma\sim T
\ee
in the quantum-critical regime, thereby allowing
one to link the kinetic and thermodynamic coefficients together \cite{0806.0110}.

Also, observe that the ratio between Eqs.(22)
\be
{\eta e^2v^2\over \sigma T^2}=const
\ee
is dimensionless and, therefore, constant in the neutral massless case. 
As such, it should be contrasted against the proposal 
$$
{\eta\over \sigma T^{2/z}}=const
$$
which was put forward for $\theta=0$ in Ref.\cite{1008.2944}.
Their obvious discrepancy (even in this limit) stems from  
the improper account of the velocity's dimension in Ref.\cite{1008.2944}. 

In the charged case ($\mu\gg T$), by using $\chi_c=dn/d\mu$
and Eq.(10) one can cast the conductivity  
in the form 
$$
\sigma\sim {e^2v^2n\over \mu\Gamma}
$$ 
Should the rate $\Gamma$ then happen to be linear, as in Eq.(27), 
the conductivity would exhibit the ubiquitous in strongly correlated systems 
$\sim 1/T$ behavior \cite{1311.2451}, seemingly in agreement
with various scenarios of the cuprates and other 'strange metals'  
that emphasize their proximity to one or another putative quantum-critical point.

The above scaling relations 
can also be generalized to include anisotropic spatial geometries.
In the simplifying case of a unidirectional rotationally anisotropic  metric \cite{1011.3117}
\be 
g_{tt}\sim -r^{2\theta/d-2z},~~ g_{\parallel,\parallel}\sim r^{2\theta/d-2w},~~g_{rr}=g_{\perp,\perp}\sim r^{2\theta/d-2}
\ee 
the scaling dimensions read
$$
[t]=-z,~~[x_{\parallel}]=-1,~~[x_{\perp}]=-w
$$
$$
[A_{\parallel}]=w-2\theta/d,~~[A_{\perp}]=1-2\theta/d,
$$
$$
[{\cal E}_{\parallel}]=z+w-2\theta/d,~~~[{\cal E}_{\perp}]=1+z-2\theta/d,
$$
$$
[J_{\parallel}]=d+z-1-\theta, ~~~[J_{\perp}]=d+z+w-2-\theta,
$$
\be
[n]=d+w-1-\theta-2\theta/d,~~~[B]=1+w-2\theta/d
\ee
where $B$ is a magnetic field perpendicular to both $\vec E$ and $\vec J$.

Choosing the axes $x$ and $y$ along the $\parallel$ and 
one of the $d-1$ $\perp$ directions, respectively, one obtains 
\be
[\sigma_{xx}]=d-1-w-\theta+2\theta/d,
\ee 
\be
[\sigma_{xy,yx}]=d-2-\theta+2\theta/d,
\ee
\be 
[\sigma_{yy}]=d+w-3-\theta+2\theta/d
\ee
In the charged case, all the components are expected to be proportional 
to the density, as the Hall response of a particle-hole symmetric system vanishes identically.
However, the common density factors cancel out in the Hall angle
\be
\cot\theta_H={\sigma_{xx}\over \sigma_{xy}}\sim {en\over B\sigma_{xx}}\sim
B^{-1}T^{(2/z)(1-\theta/d)}
\ee
where the linear proportionality of $\sigma_{xy}$ 
to a weak magnetic field has also been taken into account. 

Eqs.(36)-(39) agree with the results of Ref.\cite{1011.3117}
where only the case of $\theta=0$ was considered.  
A further generalization to the fully anisotropic case would be quite straightforward, too.

It was concluded in Refs.\cite{1011.3117} that
both goals of reproducing the linear resistivity $and$ quadratic Hall angle
characteristic of, e.g., the behavior found in the 
superconducting cuprates can not 
be achieved simultaneously, regardless of the choice of $z$ and $w$.

For instance, by choosing $z=1, w=1/2,\theta=0$ 
one does obtain $\sigma_{xx}\sim 1/T, \cot\theta_H\sim T^2$ in the charged case, although it  can only come at the expense of acquiring a 
strong spatial anisotropy (now $\sigma_{yy}\sim 1/T^2$, independent of $w$
or $d$). 

One can also check that having yet another available parameter $\theta$   
does not change the above conclusions. For instance, in 
the isotropic charged system ($z=w=1$) one can get 
$\sigma_{xx,yy}\sim 1/T$ by simply choosing $\theta=d/2$, but then the concomitant
Hall angle is $\cot\theta_H\sim T$.

It is also instructive to compare the above scaling dimensions 
to the predictions of the holographic 'membrane paradigm' which offers simple integral expressions for such important thermodynamic characteristics as charge susceptibility \cite{0809.3808}
\be
e^2\chi={\big(}\int^\infty_{r_h}dr
{\sqrt {g_{rr}|g_{tt}|\over \prod_ig_{ii}}}{\big)}^{-1}
\ee
or enthalpy density 
\be
E+P={\big(}\int^\infty_{r_h}{dr\over g_{xx}}
{\sqrt {g_{rr}|g_{tt}|\over \prod_ig_{ii}}}{\big)}^{-1}
\ee
For the HV geometry (6) Eq.(40) yields
\be
e^2\chi\sim T^{(d-\theta-z+2\theta/d)/z}
\ee
which fully agrees with (14) and (24).

In contrast, the result of computing Eq.(41)
\be
E+P\sim T^{(d-\theta-z+2)/z}
\ee
is clearly at odds with Eq.(21)  
even for $\theta=0$, as long as $z\neq 1$. 

One can readily check that for any $\theta$
such discrepancy can not be fixed by adding any 
powers of the velocity and, if taken at its face value,  
questions the validity of Eq.(41).

Moreover, the 'membrane paradigm' approach
offers a closed expression 
for the WF ratio \cite{1008.2944}
\be
{e^2\kappa\over T\sigma}={\big(}{E+P\over Tn}{\big)}^2
\ee
Physically, the WF ratio provides a measure of the energy dependence of 
the dominant scattering rate. 
The classic WF law stating a constancy of this ratio would be expected to hold
in any regime dominated by (quasi)elastic scattering, including, e.g., the 
case of electron-phonon scattering either well below or 
well above the Debye temperature.  

In the important case of a 
zero-density 'relativistic' system with $z=1$, the thermal conductivity 
appears to be formally proportional to the momentum and, therefore, 
becomes infinite in the absence of 
momentum relaxation, thus making the WF ratio diverge, in accordance with Eq.(44)
for $n=0$, and signalling an extreme form of the WF law's violation. 

However, if the density appearing in Eq.(44) were interpreted as its zero-temperature value $n(T=0)$, then in the neutral case 
the WF ratio would be divergent regardless of the value of $z$.
Conversely, if it were treated as the equilibrium $T$-dependent density of particles/holes with the dimension given by Eq.(13), then
in the neutral system the r.h.s. of Eq.(44) would always be finite, including the case of $z=1$. Clearly, a further clarification on the conditions under which Eq.(44)
holds is warranted here.

The scaling analysis of the quantum-critical regime can also be extended to spin dynamics.   
A small field expansion of the free energy yields the dimension
of the spin susceptibility
\be
[\chi_s]=[E/B^2]=d-\theta+z-4+2\theta/d
\ee
However, 
there seems to be neither a solid holographic result to compare with,
nor even a commonly accepted recipe for computing this quantity.

The few previous attempts range from using a radial equation for the variation of the vector potential $\delta A_{\perp}$ similar 
to Eq.(46) in the next Section \cite{1207.4643} 
$$
\partial^2_r\delta A_{\perp}+\partial_r\log
{(\prod_ig_{ii}|g_{tt}|)^{1/2}\over g^{1/2}_{rr}g_{xx}}
\partial_r\delta A_{\perp}
$$
$$
-(\omega^2{g_{rr}\over g_{tt}}+k^2{g_{rr}\over g_{yy}})\delta A_{\perp}=0
$$
or for the magnetic field itself $\delta B=\partial_{\parallel}\delta A_{\perp}$ \cite{1205.3267} to that for the spin connection $\delta\omega_t^{xy}\sim\partial_x\delta g_{ty}$ \cite{1304.3126}
from which one can evaluate $\chi_s$, as if it was just another 
response function of the Kubo type.

In particular, in the $2d$ case and for $\omega/T\gg 1$ 
the thus-obtained result \cite{1205.3267}, 
$
\chi_s\sim\omega^{2/3}
$,
was claimed to compare favorably with that experimentally observed 
in the conjectured spin-liquid state of the quasi-$2d$ materials $YbRh_2(Si_{1-x}Ge_{x})_2$
and $ZnCu_3(OH)_6Cl_2$. 
It can be easily checked, however, that the scaling dimension (45) does not appear to support the above estimate for any relevant values of $z$ and $\theta$ (see below).

{\it Nifty shades of holographic conductivity}

Electrical conductivity 
has been computed for a variety of holographic models and 
in a number of different ways. 
However, establishing a consistency between 
the results of different calculations (or a lack thereof) does not 
seem to have always been particularly high of the agenda.

The most frequently employed calculation of the electrical conductivity and other kinetic coefficients
is based on the holographic adaptation of the Kubo formula \cite{AdS}.
It proceeds by solving linearized equations for small variations of the electromagnetic potential $\delta A_\mu$ and (possibly) such coupled component(s) of the metric as $\delta g_{tx}$, and/or other degrees of freedom, depending on the field content of the bulk theory in question.

In the case of a generic electro-magneto-gravitational background 
treated in the customary probe limit by virtue of the
DBI action (3), the relevant quasi-normal mode obeys the equation   
\cite{1306.3816} 
\bea
\partial^2_r\delta A_i+\partial_r\log{(\prod_ig_{ii})^{1/2}|g_{tt}|\over g^{1/2}_{rr}g^{3/2}_{xx}h(k^2h^2-\omega^2)}\partial_r\delta A_i-
\nonumber\\
-{g_{rr}\over g_{tt}}(\omega^2-k^2h^2)\delta A_i=0
\eea
where
$$ 
h={\sqrt {|g_{tt}|g_{rr}-F^2_{tr}\over g_{rr}g_{xx}}}=
{\sqrt {|g_{tt}|\over g_{xx}(1+(en)^2/\prod_ig_{ii})}}
$$
As per the standard holographic 
prescription \cite{AdS}, the optical conductivity is then defined 
as the reflection coefficient of a radial in-falling wave  
\be 
\sigma=Im{r\partial_r\delta A_i\over \delta A_i}|_{r\to 0}
\ee
where the Fourier-transformed function $\delta A_i(r,\omega,k)$ and its derivative are evaluated at the boundary (a.k.a., the UV limit). 

As one technicality, in order to compute Eq.(47)  
one first solves Eq.(46) in the opposite, IR or $r\to\infty$, limit  
(which is also formally attainable by putting $n=0$) where it reads
\be
{\partial^2_r\delta A_i}+(2\gamma-\alpha+\beta){\partial_r\delta A_i\over r}
+\omega^2r^{2(\alpha-\beta)}\delta A_i=0
\ee
the coefficients being those of the metric (4).

Next, imposing the in-falling boundary condition at $r\to\infty$ one obtains   
the solution 
$$
\delta A_i\sim u^{\nu}H^{(1)}_{\nu}(u)
$$
where $\nu=1/2-{\gamma/(1+\alpha-\beta)}$ and $u=\omega r^{z}/z$,
which then has to be matched with that of 
the equation obtained from (46) in 
the $r\to 0$ limit by comparing the two in the region
$u\sim 1$ where they overlap 
(which, in turn, requires one to take the small $\omega$ limit). 

Skipping the algebra (unlike much-needed physical discussions of the results of this calculation, such formal manipulations, complete with all the auxiliary 
technical details, can be readily found in many of the pertinent papers), 
one obtains
\be 
\sigma_{Kubo}\sim \omega^{-2\gamma/(1+\alpha-\beta)}
\ee
Somewhat surprisingly, instead of deriving this general result once and for all,
in much of the holographic literature
this calculation would be performed anew for every equation of the type (48).
 
Also, observe that Eq.(49) is invariant 
under the transformation (5), in agreement 
with the aforementioned conformal equivalence 
of the corresponding metrics.

Barring the fact that under the aforementioned matching
condition the power-law dependence (49) 
is derived for low frequencies, this asymptotic behavior 
and its analogues (see below) have  
been contrasted against the experimental data
taken at energies up to $eV$ 
(e.g., in the case of the cuprate superconductors \cite{basov}).

Applying Eq.(49) to the HV metric (6) paired with the    
radial electric field (11), one obtains the optical 
conductivity of a charged ($n\neq 0$) holographic system
\be 
\sigma_{Kubo,charged}\sim \omega^{-(2/z)(1-\theta/d)}
\ee
for small $\omega$ and $z>2(1-\theta/d)$, while in the opposite case one gets 
$\sigma\sim 1/\omega$.
This result was reported in Refs.\cite{1306.3816} 
(the first two of these references addressed only the 
limit $z\to\infty, -\theta/z=const$, though).

In the neutral case, the conductivity can be obtained
by expanding and solving Eq.(46) directly at the boundary ($r\to 0$) where
the electric field is negligible, thereby yielding 
\be 
\sigma_{Kubo,neutral}\sim \omega^{(d-2)(1-\theta/d)/z}
\ee
which estimate is in a perfect agreement with the scaling dimension (25). 

In turn, Eq.(50) can then be readily rationalized by observing that the ratio  
$\sigma_{Kubo,neutral}/\sigma_{Kubo,charged}$
scales as the charge density (15), which   
in the neutral system is played by 
the density of thermally activated quasiparticles 
(of either charge sign).

Notably, in the $2d$ case Eq.(51) allows for no faster than logarithmic dependence.
Besides, there seems to be little difference between the general HV and the Lifshitz ($\theta=0$) geometries. 
In that regard, it is worth mentioning that 
the experimentally measured optical conductivity of a neutral (undoped) $2d$ Dirac metal, such as graphene, indeed appears to be nearly constant ($\sim e^2/h$). 

Specifically, in Ref.\cite{0903.4178} the latter was found to behave as
\be
\sigma_{graphene}\sim Re{T\over i\omega+g^2(T)T}
\ee
where the logarithmically running effective charge 
$g(T)\sim 1/\log T$ represents the effect of the Coulomb interactions. 

In the higher dimensions $d>2$, according to 
Eq.(51) the conductivity of a neutral 
system generally vanishes for $\omega\to 0$, regardless of the value of $z$,
which behavior is consistent with the intrinsically semi-metallic
nature of such systems.  
In the pertinent example of the $3d$ 'Weyl metal' where $z=1$ it was recently found that  $\Delta_{\sigma}=3-4M$ with $|M|<1/2$ \cite{1403.3608}.

It is again instructive to compare Eqs.(50,51) to the predictions of 
the 'membrane paradigm' which also provides a simple algebraic expression for
the low-$\omega$ value of the conductivity.  
The latter is cast solely in terms of the geometry at the (necessarily, non-degenerate) horizon (thus, such results would not be applicable to the extremal black branes) without the need of solving any differential equations. 

Furthermore, this approach can also be extended
to include a magnetic 
field. To first order in the weak field $B$, both the diagonal and off-diagonal components of the DC conductivity tensor take the following closed form \cite{1011.3117}
\bea 
\sigma_{xx}\sim e^{-2\phi_0}
{{\sqrt {(en)^2+e^{2\phi_0}\prod_i g_{ii}}}\over g_{xx}}|_{r\to r_h} \nonumber\\
\sigma_{xy}\sim {e^{-4\phi_0} enB\over g_{xx}g_{yy}}|_{r\to r_h}
\eea
where $\phi_0$ is the fixed point value of the dilaton. 

Quite remarkably, for $\phi_0=0$ and $n\neq 0,~~~\phi_0=0$
the first of Eqs.(53) exactly reproduces Eq.(50) and (51)
for $n\neq 0$ and $n=0$, respectively,
whereas the second one is fully consistent with the scaling dimension (37). 

The problem, however, is that, as opposed to Eqs.(50) and (51), 
Eqs.(53) are supposed to be evaluated at the horizon, rather than the boundary.

In general, the local conductivity 
defined according to Eq.(49) at an arbitrary $r$,
is expected to be independent of $r$, unless
there is a 'running' field, such as an electrical scalar potential or
dilaton which can bring about a non-trivial $r$-dependence.

In the neutral case and in the absence of a dilaton, 
no radial evolution should indeed occur, and so the conductivity 
could be equally well evaluated either at the boundary or the horizon
($\sigma^{(B)}_{Kubo,neutral}=\sigma^{(H)}_{Kubo,neutral}$). 

However,  
in the charged case one would expect the local conductivity
to vary with the radial variable, 
whether or not a non-trivial dilaton field is present. 

To that end, in Ref.\cite{1008.2944} a general relation was proposed 
\be
{\sigma^{(B)}\over \sigma^{(H)}}={\big (}{ST\over E+P}{\big )}^2
\ee 
which ratio becomes unity at zero density, as per the equation of state (10).  

In contrast, at finite density Eq.(54) implies a non-trivial 
radial (hence, temperature) dependence of 
the local conductivity, thereby predicting the low-$T$ behavior 
$$
{\sigma^{(H)}_{Kubo,charged}\approx\sigma^{(B)}_{Kubo,charged}}
{\big (}{n\mu\over ST}{\big )}^2\sim T^{(2/z)(-d+\theta-z-1+3\theta/d)}
$$
which clearly contradicts Eq.(50).

Conversely, if one chooses to treat Eq.(50) as the horizon value 
$\sigma^{(H)}_{Kubo,charged}$ then Eq.(56) implies 
$$
\sigma^{(B)}_{Kubo,charged}\sim T^{(2/z)(d-\theta+z-1-\theta/d)}
$$ 
which is again different from the predictions of the scaling analysis.

This adds to the argument that a better understanding of the applicability 
of such formulae as Eqs.(44) and (54) for $n\neq 0$ is definitely called for.

However, in spite of some confusion with their terms of use, 
Eqs.(53) can still capture such intrinsic properties of the conductivity tensor as, e.g., the relative scaling of its components with temperature.  

Namely, the scaling dimensions (36-38) 
would seemingly imply that the following 
relation 
\be
\sigma_{xy}\sim B{\sigma^2_{xx}\over en}
\ee 
sets in, as the system approaches the neutral regime
at high temperatures. In fact, such a relation does hold - but only for the partial
(particle and hole) contributions towards the total Hall conductivity.
Should both components happen to have equal mobilities (as, e.g., in the case of a  
particle-hole symmetric spectrum), the overall $\sigma_{xy}$ would only be  
proportional to the charge imbalance given by $en(T=0)$, 
thereby resulting in the different relative scaling rule
\be
\sigma_{xy}\sim Ben(T=0){\sigma^2_{xx}\over (en)^2}
\ee
The naive relation (55) could still hold, though, if the 
spectrum were lacking particle-hole symmetry (as, e.g., in the case of topological insulators where the Dirac spectrum emerges as a result of the bulk gap inversion).

In that regard, the recent Ref.\cite{Karch} claimed that in the extended
class of the $3$-parameter HV systems and at sufficiently high temperatures
the cuprate-like behaviors of, both, $\sigma_{xx}$
and $\sigma_{xy}$ could be recovered even in the spatially isotropic case. 

Indeed, by using Eq.(56), one can see that 
in the original, $2$-parameter HV system with $z\neq 0$ and $\theta\neq d$,
this does not happen, as the desired dependencies $\sigma_{xx}\sim T^{-1}$ and $\sigma_{xy}\sim T^{-3}$ 
would only occur in the unphysical dimension $d_s=2/3$.

However, in the $3$-parameter family of 
the generalized HV systems, such dependencies
could emerge under the choice of parameters \cite{Karch}
$$
\phi=2-3z/2,~~~\theta_m=d_s-z/2
$$
which conditions would only be consistent with Eq.(27), provided that 
$$
d+2-2z-\theta=0
$$
(cf. Eq.(5.5) in Ref.\cite{Karch} where $d$ is replaced with $d_s$ - obviously, in error,
as its derivation utilizes Eq.(3.16) where the identification $d_s=d$ has already been made).  

The physical tangibility of such a scenario remains to be discerned, as does the 
whole notion of different dimensions for the gravitational and gauge (matter) degrees of freedom. In the known examples of layered strongly correlated systems, the emergent 
(both, gauge and matter) 
fields always tend to be confined to the supporting lower-dimensional subspace.
On the other hand, in, e.g., graphene the Coulomb interactions do permeate the surrounding $3d$ space, but then their dynamics becomes affected by the charges outside
the graphene plane, thus hindering the possibility of finding a closed holographic description.  
 
One more remark is in order here.
According to Eqs.(53), a finite longitudinal (Ohm's) conductivity can arise due to, both, 
current relaxation as well as (Schwinger's) pair production.
While the latter mechanism operates even at 
zero charge density (unlike the current, the system's 
momentum can then remain conserved), the former one 
requires a finite density of carriers immersed
in a dissipative medium composed of some neutral modes.

Moreover, in the framework of the 'membrane paradigm' 
the two sources of finite conductivity combine together in a rather peculiar manner 
\be
{\sigma}={\sqrt {\sigma_{mr}^2+\sigma_{pc}^2}}
\ee
where $\sigma_{mr}$ and $\sigma_{pc}$ 
stand for the contributions due to
momentum relaxation and pair creation, respectively.

For comparison, yet another recent application of the Kubo 
approach yields \cite{1401.5436}
\be
{\sigma}=\sigma_{mr}+\sigma_{pc}
\ee

Notably, both Eqs.(57) and (58) violate the standard Matthiessen's rule,
according to which it is the $\it inverse$ of the partial 
conductivities, rather than the conductivities 
themselves, that tend to add up.
Instead, Eq.(58) is reminiscent of the combination rule for those scenarios
where more than one type of current carriers 
(as opposed to more than one mechanism
of scattering for the same type of carriers) is present, and the best conducting one short-circuits the rest of the system. 

Moreover, the two contributions entering Eqs.(57),(58) were found to behave as
\be
\sigma_{mr,1}(\omega)\sim \omega^{|3+(d-2-\theta)/z|-1}
\ee
\cite{1308.2084} or 
\be
\sigma_{mr,2}(\omega)\sim \omega^{|1+(\theta-2-d)/z|-1}
\ee
\cite{1401.5436}, whereas
\be
\sigma_{pc}(\omega)\sim \omega^{(|1-\zeta|-1)/z}
\ee
\cite{1308.2084,1401.5436} where a new ('conduction') exponent $\zeta$ 
was introduced to describe the scalar potential $A_0\sim r^{\zeta-z}$.
Conceptually, it can be related to the aforementioned $\phi$-factor.
 
Given that the exponents appearing in Eqs.(59)-(61) differ from those 
discussed earlier in this Section, 
a better understanding of their  
physical nature as well as the origin of the combination rules (57) and (58),
would once again be warranted.

Interestingly, though, the momentum relaxation exponent (60) found in Ref.\cite{1401.5436}
agrees with the results of still another recent work of Refs.\cite{1401.7993} 
where an important effort was made to include elastic scattering, 
alongside the inelastic one.  

In fact, the analysis of Refs.\cite{1401.7993} represents  
a 'holography-augmented' transport theory, 
rather than a systematic all-holographic calculation.  
Conceivably, though, such a hybrid approach might be better 
equipped for capturing the underlying 
physics of the relevant transport phenomena.

Specifically, these works employed the so-called memory function formalism 
\cite{1201.3917}
which does not explicitly rely on the existence of 
well-defined quasiparticles and presents, e.g., the electrical 
conductivity in the form
\be 
\sigma_{memory}=\chi_{JP}^2 (\int dkk_x^2
{Im D(\omega,k)\over \omega}-i\omega\chi_{PP})^{-1}
\ee
where $\chi_{JP,PP}(T)$ are the current-momentum and momentum-momentum susceptibilities.

The formula (62) assumes that momentum is the only (nearly) conserved physical quantity
and relates the conductivity to the spectral density of the operator that breaks momentum conservation. It is expected to work best in the hydrodynamic regime where the rate of
momentum relaxation due to a breaking of translational invariance by elastic impurity or lattice-assisted inelastic Umklapp  
scattering is smaller than the inelastic rate which controls 
a formation of the hydrodynamic state itself. 
For instance, in the case of $\mu\gg T$  
the rate of the Umklapp scattering is of order $\sim T^2/\mu$, whereas the latter one 
is given by the universal quantum-critical rate (32).
 
In general, the onset of hydrodynamics is a distinct property of strong correlations
which would be routinely absent in the Fermi liquid regime.
Such a regime would also be absent in $1d$, thanks to the peculiar $1d$ kinematics facilitating the emergence of infinitely many (almost) conserved currents. 

In the absence of any (nearly) conserved quantities Eq.(62) ceases to be applicable.
Although the corresponding 'incoherent' metals do not allow 
for any simple description, they have been eloquently argued  \cite{1405.3651}
to conform to the ubiquitous $\sigma\sim 1/T$ dependence stemming from 
the universal scattering rate (32). 

In many cases, though, strong interactions often go head-in-glove with (and enhance the effects of) strong disorder.
The combined effects of the two can hardly be accounted for by means of the 
perturbative Altshuler-Aronov theory and are likely to require
some intrinsically non-perturbative approaches, such as the Efros-Shklovskii
one, thus allowing for other, essentially non-linear, $T$-dependencies,
$\sigma\sim \exp(-{\#}/T^{\alpha})$.

It is also worth noting 
that, unlike Eqs.(57) and (58), the applications of Eq.(62) 
would have a good chance to be in compliance with the Matthiessen's  rule, as the different scattering mechanisms tend to correspond to separate contributions to the integral kernel $D(\omega,k)$, thereby producing additive terms in the expression for the
$inverse$ conductivity. 

The main inference from Eq.(62) is a transfer of the spectral weight
from the coherent Drude peak to the incoherent high-frequency tail. 
It is worth noting, though, that in the previous applications of Eq.(62)
a possible quasiparticle renormalization  
was not, $de~facto$, considered, 
as the behavior of $\chi_{JP,PP}$ was believed to be non-singular 
and, at most, only weakly $T$-dependent (such an assumption notwithstanding, e.g.,
at the onset of the Mott transition one expects $\chi_{JP}=0$).

As mentioned above, the behavior found in Ref.\cite{1401.7993} 
\be
\sigma_{memory}\sim T^{(\theta-2-d)/z}
\ee
coincides with that reported in Ref.\cite{1401.5436}. However, in Ref.\cite{1401.7993}
it was shown to emerge only in the strong coupling regime, whereas
the lowest (second) order perturbative result was found to 
be non-universal 
$
\sigma_{memory}\sim T^{(z-d-\delta)/z}
$, 
$\delta$ being the anomalous dimension of the operator 
that breaks momentum conservation.

Taking into account the HV scaling relations (16) one observes that in the neutral case
Eq.(63) appears to be inversely proportional 
to entropy (equivalently, specific heat or viscosity), 
as conjectured earlier in Ref.\cite{1311.2451}.
However, the dependence $\sigma\sim 1/S_{ent}\sim T^{(\theta-d)/z}$ 
advocated in \cite{1311.2451} can only occur in the limit $z\to\infty$. 
Otherwise, Eq.(63) features an additional factor that, incidentally,
behaves as the inverse square of a $T$-dependent 'graviton mass', $m\sim T^{1/z}$.

In a series of works \cite{1306.5792}, 
it was indeed proposed to incorporate the effects of static disorder
by introducing a graviton mass $m$ which is
weakly (if at all) $T$-dependent. Although under such an assumption
the desired dependence $\sigma\sim 1/S_{ent}$ does indeed set in, it remains to be seen 
whether such a scenario can be justified beyond the $ad~hoc$ level. 

As yet another effort towards marrying 
the formal holographic manipulations with the more traditional transport theory, it was also proposed to mimic the momentum-relaxing Umklapp processes 
brought about by the presence of a regular crystal lattice with 
expressly anisotropic geometries and periodic scalar and/or dilaton
potentials.

To that end, in the previously quoted  Ref.\cite{1204.0519} 
the crystal lattice was modelled by a periodic 
electric potential, resulting in 
$\Delta_{\sigma}={-2/3,-{\sqrt 3}/2}$ in $2d$ and $3d$, respectively.

Also, noteworthy is the proposal \cite{1212.2998}
to use the helical Bianchi-$VII_0$ metric with a pitch in the $x$-direction 
\bea 
g_{tt}=-g_{rr}\sim -1/r^{2},~~~g_{xx}\sim r^{2/3},\nonumber\\
g_{yy}\sim 1/r^{4/3}, g_{zz}\sim 1/r^{2/3}
\eea 
as a holographic description of the anisotropic $3d$ periodic structure which  
gives rise to the interaction-induced Mott-type state with the 'bad-metallic'  
conductivity $\sigma_{xx}\sim T^{4/3}$ in the direction of the pitch, alongside a gapless behavior of the entropy, $S_{ent}\sim T^{2/3}$.

Although the Bianchi geometry (64) does have its intellectual appeal,
it should be noted that the above choice is not unique.
As follows from Eq.(51), the same behavior of, both, $\sigma_{xx}$ and entanglement
entropy  can be found for an entire family of the uniaxially anisotropic $3d$ metrics 
\be
g_{tt}\sim -1/r^{2\alpha},~~g_{rr}\sim 1/r^{2\beta},~~~g_{ii}\sim 1/r^{2\gamma_i}
\ee
which satisfy the conditions $2\gamma_x/(1+\alpha-\beta)=-2/3,~~~\sum_i\gamma_i=2/3$. 
For instance, by choosing $\alpha=\beta=1,~~\gamma_x=-\gamma_y=-\gamma_z=-2/3$
one finds that fully spatially anisotropic geometries, as in (64), 
may not be necessary for constructing a holographic dual of 
the 'bad metal', after all.

Continuing with the list of the previously obtained holographic results 
it might be worth mentioning a few more examples
whose physical interpretation (as well as mutual consistency)
is yet to be ascertained.

For one, there has been a variety of predictions for 
the dimension of electrical conductivity.
In the neutral Lifshitz case ($\theta=0$),  
Ref.\cite{1304.7481} found $\Delta_{\sigma}=(d+2-2z)/z$,
while Ref.\cite{0607522} reported $\Delta_{\sigma}=(3-d-z)/z$,
and Ref.\cite{1306.1517}
arrived at the exponent $\Delta_{\sigma}=(d+2z-4)/z$.

None of these values appears to   
be consistent with the above scaling predictions and the
universal quantum-critical scattering rate (32).

Going beyond the Lifshitz case, Ref.\cite{1208.2535} found $\Delta_{\sigma}=3$
for $d=3$ and $\Delta_{\sigma}=(2z-3)/z$ for $\theta=d-1$,
whereas Ref.\cite{1209.3946} reported 
$\Delta_{\sigma}=(d-2)/z$, but only for $z=d-2$, 
$\Delta_{\sigma}=1/z$ for $z=(d-4)/3$, and 
$\Delta_{\sigma}={(d-\theta)/z}$, also in 
conflict with the above scaling results.

Also, for $d=2$ Ref.\cite{0911.3586} delivered $\Delta_{\sigma}={2}$, while
Ref.\cite{0807.1737} obtained $\Delta_{\sigma}={7/2}$.
Moreover, Ref.\cite{1107.2116} presented an even greater variety of values, 
$\Delta_{\sigma}=1, 2, 3, d, d-2, d-4$ for $z=1$
and $(2z+d-2)/z$ for $z\neq 1$, as well as a whole discrete
series $(1+3p)/(3+p)$, whereas other works featured 
the entire plethoras of non-universal exponents as functions of one 
or even two  continuous parameters appearing in
the holographic Lagrangian \cite{charmousis}. 

On the other hand, Ref.\cite{1107.5822} utilized the metrics (4)
with $\beta=2-\alpha$ obtaining the results
$\Delta_{\sigma}=-2\gamma/(2\alpha-1)$ and
$\cot\theta_{H}\sim T^{2\gamma/(2\alpha-1)}$, in agreement with (49).

Still other available methods of computing conductivity include extracting it from the hydrodynamic expansion or computing a drag force for massive charge carriers.
Although some of those results may seem more plausible than others,
they are still awaiting for their physical interpretation
and a systematic comparison with the predictions made by the alternative techniques.

In that regard, the general universal relations,  
such as Eq.(28) or (33), provide an important consistency test, while reinforcing 
the notion that the dynamic properties of quantum-critical systems are 
closely related to their thermodynamics.  Technically, such a relationship implies
that, apart from the relaxation rate (32), the kinetic coefficients can be found in terms of the thermodynamic ones.

Yet another important test would be provided by the sum rules for the optical conductivity
and other kinetic coefficients, akin those extensively employed in Ref.\cite{1206.3309}. 
Obviously, no monotonic low-frequency asymptotic obtained by solving 
the differential equation (46) in the Kubo formula approach  
can be up for this test. 
However, the frequency-dependent counterparts of the purely algebraic Eqs.(53) could indeed 
be used to that effect, once their closed expressions are 
obtained in a wide range of frequencies.

{\it Mother of all non-Fermi liquids}

One concrete context for a comparative 
discussion of the different holographic models 
is provided the theories of fermions coupled to gapless over-damped bosonic modes. This 'mother of all NFLs' has long been at the forefront of  theoretical research, since 
the singular interactions mediated by soft gauge field-like bosons are often associated with incipient ground state instabilities and concomitant NFL types of behavior. 

Such effective long-range and strongly retarded interactions may occur even in microscopic systems with purely short-ranged couplings. In the close proximity to a quantum-critical point, the role of the corresponding modes is then played by (nearly) gapless excitations of an emergent order parameter. 

Important examples include such problems as ordinary electromagnetic fluctuations in metals and plasmas, spin and charge ordering transitions in itinerant (anti)ferromagnets, compressible Quantum Hall Effect, Pomeranchuk instabilities resulting in rotationally anisotropic 'quantum nematic' states, etc.

Despite the differences in their physical nature, all these systems conform to the general problem of a finite density fermion gas coupled to an overdamped bosonic mode whose own dynamics is governed by 
the (transverse) gauge field-like propagator
\be
D(\omega,q)={1\over |\omega|/q^\xi+q^\rho}
\ee
In the context of electrodynamics of conducting media, 
the first and second terms
account for the phenomena of Landau damping and diamagnetism, respectively.

Over the past two decades this problem has been repeatedly attacked with a variety of techniques.

At the early stage, it was believed that the functional form of
the one-loop fermion self-energy
\be
\Sigma(\omega)=\int d\epsilon d^dq{D(\omega,q)\over (i\omega+i\epsilon-{\bf v}{\bf q})}
\sim\omega^{d-1+\xi\over \xi+\rho} 
\ee
survives in the higher orders of perturbation theory, akin to the situation in the Eliashberg theory of electron-phonon interactions \cite{aim}. 
However, the more recent analyses demonstrated an inapplicability of the naive weak coupling and $1/N$-expansions \cite{gauge}, thus calling the earlier results into question. 

There have been also attempts to study this theory without introducing the Landau damping from the outset \cite{1307.0004}, which analysis yields a self-energy $\Sigma\sim\omega^{1-\epsilon/4}$ in $d=3-\epsilon$ dimensions  and for $\rho=2,~\xi=1$ 
that is markedly different from the counterpart of (67), $\Sigma\sim\omega^{1-\epsilon/3}$.
In still other approaches, the problem was attacked by expanding in $z_b-2$ \cite{mross} or $d=5/2-\epsilon$ \cite{SSLee}.

Despite somewhat conflicting results, Eqs.(66) and (67)  
would often be used for evaluating the boson and fermion dynamical
exponents
\be
z_b=\rho+\xi,~~~z_f={\xi+\rho\over d-1+\xi}
\ee

Conceivably, a hypothetical holographic dual (if any) of the boundary theory with the interaction (66) might involve such bulk degrees of freedom as 
gauge potential, metric and scalar fields and, therefore,
it could be envisioned amongst the solutions of the EMD Lagrangian (9).

Along these lines, in Ref.\cite{dvk} a comparison was made between 
the two-point correlation function computed 
holographically in a yet-to-be-specified HV geometry 
and those obtained directly in the boundary 
theory with the use of the eikonal technique.

The agreement was found, provided that the $\theta$-parameter of 
the HV metric (6) was chosen as 
\be
\theta=d{\rho+1-d\over \xi+d-1}
\ee
thereby satisfying the relation $z_f=1+\theta/d$ and, incidentally, turning the first 
of the conditions (8) into an exact equality.
In particular, for $d=2,~~\rho=2,~~\xi=1$ 
one obtains $z_f=3/2$ and $\theta=1$ which values have also
been independently singled out on the basis of 
analysing the entanglement entropy \cite{1308.3234}.

In Table I, we compare the exponents governing a power-law decay of the conductivity computed holographically with the use of Eqs.(50),(51), and (62). These values pertain to 
the aforementioned models I-IV and are complemented by those for 
the new model V which is characterized 
by the exponents $z_f$ and $\theta$ given by Eqs.(68) and (69), respectively.
  
The first two columns contain the exponents 
$\Delta^{(\infty)}_{\sigma}$ governing the $\omega$-dependence 
for $\omega\gg T$ and given by Eqs.(50) and (51), whereas 
the third column contains the values of $\Delta^{(0)}_{\sigma}$
pertinent to the $T$-dependence for $\omega\ll T$ and  given by Eq.(62)
(a potentially strong $\omega$-dependence of the functions $\chi_{JP,PP}$ complicates the analysis of $\sigma(\omega)$ in the framework of the memory function method). 

\begin{tabular}{l*{6}{c}r}
$\Delta_{\sigma}$ & $Kubo_{charged}$ & $Kubo_{neutral}$ & Memory function \\
\hline
Model I & ${-2\over dz_f}$ & ${d-2\over z_fd}$ &  ${-3\over z_f}$  \\ 
Model II & ${-2\over z_b}$ & ${d-2\over z_b}$ & ${-(d+2)\over z_b}$ \\
Model III & ${-2\over z_b}$ & ${d-2\over z_b}$ & $-{d\over z_b}-{2\over z_f}$  \\
Model IV  & ${-2\over z_fd}$  & ${d-2\over z_fd}$  & $-{1\over z_f}-{2\over z_b}$ \\
Model V   & $2-{4\over z_f}$  & $2-d+{2(d-2)\over z_f}$ & $d-{2(d+1)\over z_f}$  \\
\end{tabular}

A few comments are in order: 

Firstly, despite being spurious as far as its physical implications are concerned 
(see below), the much-desired exponent $-2/3$ is quite robust and can be obtained 
in any of the models I-IV for both $d=2$ and $d=3$, 
as long as $\xi=1$. 
Moreover, for $\xi=1$ all the results for the models I and III as well as those for the models II and IV are identical.

Secondly, the model V with its conjectured boundary dual represented by the gauge-fermion model can also be amenable to the application of the standard Drude theory.
The latter (nominally) assumes the existence 
of a quasiparticle description and 
yields the conductivity 
\be
\sigma_{Drude}\sim Re{1\over i\omega/Z+\Gamma_{tr}}
\ee
where the (possibly strong) quasiparticle renormalization is accounted for via
the Green function's residue, $Z=1/(1-\partial\Sigma/\partial\omega)$. 

Potentially, Eq.(70) can take rather different forms at small and large $\omega$, as compared to $T$, depending on whether or not the transport scattering rate
$\Gamma_{tr}$ behaves differently from that of quasiparticle decay.

Estimating the latter as 
$
\Sigma(\omega)\sim\omega^{1/z_f}
$
one then finds the former (as well as the entire $T$-dependent DC conductivity) 
to be governed by the modified exponent 
\be
\sigma_{Drude}^{(0)}\sim 1/\Gamma_{tr}\sim T^{-(1/z_f+2/z_b)}
\ee

In contrast, at high $\omega$ one obtains 
\be
\sigma^{(\infty)}_{Drude}\sim {Z^2\Gamma_{tr}\over \omega^2}
\sim\omega^{-1/z_f+2/z_b}
\ee

Thus, the DC and AC Drude conductivities exhibit the same exponent 
only in the limit $z_b\to\infty$
which corresponds to a particular case of the generic short-ranged (and, therefore, only weakly momentum-dependent) scattering mechanism.

By contrast, for any finite $z_b$ there will be a disparity between 
the transport and quasiparticle decay rates 
and, as a result, different values of the
exponents controlling the $\omega$ and $T$ dependencies. This fairly mundane
observation should be contrasted 
with such exotic proposals as a 'wrong' 
sign of the expression under the absolute value in Eqs.(59)-(61)
or a parameter-dependent dominance of one term in Eq.(58) over the other
which were put forward in Ref.\cite{1401.5436}.

By applying Eq.(70) to the Model V one observes that 
the low-$\omega$ Drude conductivity (specifically, its exponent
$\Delta_{\sigma}^{(0)}$) 
agrees with that of the memory function approach for, at least, one of the models I-IV,
while in the high-$\omega$ regime 
(i.e., for $\Delta^{(\infty)}_\sigma$) this is generally not the case.

Specifically, for $d=2,~\rho=2,~\xi=1$ the high-$\omega$ Drude formula yields $\Delta^{(\infty)}_{\sigma}=0$, while in $3d$ one gets $\Delta^{(\infty)}_{\sigma}=-1/3$.
By contrast, in the DC limit one gets $\Delta^{(0)}_{\sigma}=-4/3$ and $-5/3$ (up to a power of logarithm) 
in $2d$ and $3d$, respectively. 

In the case of $\rho=2, \xi=0$, the counterparts of the above values read
$\Delta^{(\infty)}_{\sigma}=1/2$ ($2d$) and $0$ in ($3d$), while 
in the DC limit one gets $\Delta^{(0)}_{\sigma}=-3/2$ and $-2$ in $2d$ and $3d$, correspondingly.

The latter estimate should not be compared directly with the 
prediction $\Delta^{(\infty)}_{\sigma}=-1/3$ made in Ref.\cite{hhms} 
for the scenario of an incipient $2d$ spin density wave instability with the large momentum. The cuprate-like shape of the Fermi surface and the dominant scattering 
involving its opposite regions modify the above results obtained 
under the assumption of a spherical Fermi surface by a missing
factor of $\omega^{2/z_b}$ due to the scattering between the conjugate
pairs of hot spots and an additional 
factor of $\omega^{1/2}$ due to a finite span of the region around 
each hot spot. Together, the two effect conspire to result 
in a somewhat accidental cancellation, thereby producing $\Delta^{(\infty)}_{\sigma}=0$.

As to the quoted exponent $-1/3$, it was obtained in \cite{hhms}
by going well beyond the Drude approximation and focusing on certain 'energy transfer' processes which involve pairs of soft bosons  with small total momenta. 

The story does not seem to end there, though. The recent calculation for the 'Ising nematic' model produced yet another term that dominates over all the other ones at low $T$ \cite{1401.7993}
\be
\sigma_{IN}\sim(T\log T)^{1/2}
\ee
which was associated with the dominant process of scattering off of 
a (quasi)static 'random Ising magnetic field'. 

Under a closer inspection, the behavior (73) turns 
out to be indicative of the IR divergence 
of the momentum integral in Eq.(62) from which it is rescued by 
introducing a cut-off at energies of order the mass of the bosonic 
mode $m\sim(T\log T)^{1/2}$. 

However, should such a mass happen to 
be prohibited on the grounds of, e.g., unbroken gauge invariance, 
the problem in question would turn out to be 
intrinsically strongly-coupled
and possibly resulting in quite different, yet to be determined, conductivity behavior
(in the case of the Ising nematic, this possibility was claimed to be conveniently pre-empted by the onset of a superconducting instability \cite{1403.3694}).

Besides, albeit being seemingly innocuous to first order \cite{1401.7993}, 
the $T$-dependent corrections  
to the susceptibilities $\chi_{JP,PP}$ may get promoted to the exponent
in the higher orders, thus altering the overall power counting.  

In that regard, a particularly interesting would be the actual gauge field problem where, unlike the longitudinal, the transverse gauge boson 
does not develop any (thermal) mass, except 
in the case of a symmetry-breaking phase transition.

{\it Real-life non-Fermi liquids}

The list of documented NFLs is extensive and includes 
ferromagnetic metals (e.g., $MnSi, ZrZn_2$) and superconductors ($UGe_2$, $URhGe$, $UCoGe$), heavy fermions (e.g., $YbRh_2Si_2$, $CeCoIn_5$ or $URu_2Si_2$), unconventional superconductors such as cuprates and iron pnictides, electronic nematics (e.g., $Sr_3Ru_2O_7$), insulating magnets (e.g., $CoNb_2O_6$ and $TlCuCl_3$), quasi-one dimensional Mott insulators (e.g., $(TMTSF)_2PF_6$ or $(TMTSF)_2ClO_4$), etc. 

Given that in most cases the dynamical exponent $z>1$, one might naively expect all the $3d$ systems to show the classical mean-field scaling behavior, since the effective dimension of spatial fluctuations,  
which equals $d+z$, appears to exceed the upper critical dimension $d_{uc}=4$.

Moreover, the FM systems with a conserved order parameter and 
$z_b=\rho+\xi=3$ would be anticipated to follow the classical 
scenario for any $d>1$, whereas for the AFM ones (where the order parameter is not conserved
and $z_b=\rho+\xi=2$) it would then happen for all $d>2$.

However, this argument can be invalidated by dangerously irrelevant variables presenting a potential source of hyperscaling violation and resulting in 
the breakdown of the corresponding relation between the specific heat ($\alpha$)
and correlation length ($\nu$) exponents, 
$d\nu=2-\alpha$ \cite{0103393}. 

For one, the $3d$ helical ferromagnet
$MnSi$ demonstrates a NFL behavior for  
$\sigma\sim T^{-3/2}$ 
(which reverts to $\sim T^2$ in a field of $3 T$). The itinerant ferromagnet 
$ZrZn_2$ shows somewhat similar properties. 
Such prototypical NFL materials have long been viewed as potential
candidates to the application of the $3d$ gauge-fermion theory discussed in the previous Section, although the thus-obtained conductivity would behave 
as $\sim T^{-5/3}$, in disagreement with the above dependencies.

In turn, the $3d$ AFM heavy-fermion compound 
$YbRh_2Si_2$ exhibits a quantum-critical point 
at a finite field $H_c$, featuring 
$\chi_s\sim T^{1/4},~~C\sim T^{3/4},~~\sigma\sim T^{-3/4}$, which behavior is suggestive of the critical exponents $z=4$, $\alpha=1/4$, and $\nu=1/3$. 

Its doped cousin $YbRh_2(Si_{1-x}Ge_{x})_2$  
shows a power-law behavior of the low-$\omega$ spin susceptibility
$Re\chi_s\sim T^{-0.6},~~Im\chi_s\sim\omega/T^{1.6}$ for $x\approx 0.05$ 
\cite{gegenwart}.

Another example of the $3d$ AFM materials, $UCu_{5-x}Pd_x$, 
manifests $C\sim T^2, ~~\sigma\sim T^{-1/3}$ and 
$\chi_s\sim T^{\gamma}$ where $\gamma$ ranges between $0$ for $x=1$ (i.e., $\chi_s\sim\ln T$) and $\chi_s\sim T^{-1/3}$ for $x=1.5$ \cite{0103393}.

The list of the $3d$ AFM also includes $CeIn_3$ with
$\sigma\sim T^{-3/2}$ (under near-critical pressure), 
$CePd_2Si_2$ with $\sigma\sim T^{-5/4}$,
$CeRnSn$ with anisotropic resistivity: 
$\sigma_{ab}\sim T^{-3/2}$, $\sigma_c\sim 1/T$, and magnetic susceptibility:
$\chi_{s,ab}\sim T^{-1/3},~~\chi_{s,c}\sim T^{-1.5}$.

Another (this time, quasi-$2d$) AFM material,  
$CeCu_{6-x}Au_{x}$, shows $C\sim T^{7/8},~~
\sigma\sim T^{-7/8},~~\chi_s\sim T^{1/8}$ for $x\approx 0.1$ 
\cite{schoeder} which data hint at the exponents $ z=8/3, \alpha=1/8, \nu=3/7$.

It was also argued in Ref.\cite{0103393} that such data 
could be explained in terms of the anisotropic dynamical susceptibility $\chi_s(\omega,
k)=(k_\perp^2+k_\parallel^4+|\omega|^\gamma)^{-1}$ where $\gamma=4/5$, 
yielding $\rho\sim T, \chi_s\sim T^{1/2}$.

As regards the quasi-$2d$ AFM materials, the Kagome AFM 
$ZnCu_3(OH)_6Cl_2$ (a.k.a. Herbertsmithite) shows a 
power-law behavior of the bulk susceptibility, 
$\chi_s\sim T^{-2/3}$, and the spin relaxation rate $1/T_1\sim T^{0.7}$,
although the issue of its possibly non-analytical $\omega$-dependence at 
small $\omega$ has not been completely settled yet \cite{helton}.

Also, the in-plane optical conductivity of this material  $\sigma\sim\omega^{7/5}$ 
was argued to be consistent with the picture of a spin gapless (since $C\sim T$)
but charge gapped $2d$ Dirac spin liquid state \cite{pilon}.

The gauge theory of the $U(1)$ spin-liquid states is also
expected to reproduce such observed metal-like properties as 
$C/T, \chi_s, \kappa/T\to const$ (despite $1/T_1\sim T^2$ which might indicate a soft nodal gap) in the organic compound $EtMe_3Sb[Pb(dmit)_2]_2$ which shows the conductivity exponent  $\Delta_{\sigma}$ varying between $3/4$ to $3/2$ \cite{yamashita}. 
A similar spin-liquid state 
(although, possibly, with a small spinon gap) characterized by the conductivity
exponent $\Delta_{\sigma}$ ranging between $0.8$ and $1.5$
occurs in $\kappa$-BEET-$Cu_2CN_3$ \cite{1208.1664}).

While the complete theory is still being developed and perfected,   
its viable variant was proposed 
in the framework of the phenomenological analysis of Ref.\cite{1303.3926} 
where the NFL self-energy was assumed to be independent of momentum, $\Sigma(\omega)\sim\omega^{1-\alpha}$. 
The latter was found as a self-consistent solution of the self-consistent equations 
\be
\Sigma(\omega,q)=\int{d\epsilon d^dp}
\Lambda^2(\epsilon)\chi_E(\omega+\epsilon,p+q)G(\epsilon,p)
\ee
where $G(\omega,q)=(i\omega/Z-{\bf v}{\bf q})^{-1}$
and 
$
\chi_E(\omega,q)=\int {d\epsilon d^dp}GGDD
$
is the effective propagator of soft bosonic pair-exchange processes. 
The interaction vertices $\Lambda$ are decorated with the renormalization factor $Z$, 
enforcing the corresponding Ward identity.

The exponents obtained by solving Eq.(74) turn out to be solely 
determined by the spatial dimension
\be 
\alpha=1/2-1/z_b,~~~\nu={1\over 2+z_b\alpha},~~~z_b=4d/3,~~z_f={1\over 1-\alpha}
\ee
The resulting observables
\be
C\sim T^{1-\alpha},~~~\sigma\sim\omega^{\alpha-1}~~~\chi_s\sim T^{\alpha}
\ee
turn out to describe quite well the aforementioned data on 
$YbRh_2Si_2$ and $CeCu_{6-x}Au_{x}$ for $d=3$ and $2$, respectively. 

It would be a real challenge (and an impressive achievement in the case of success) for the holographic approach to reproduce more than one of the above exponents 
(for the same material). 

{\it Summary}

The holographic approach aspires to provide a potential 
framework for treating those strongly coupled systems that do not fit into the conventional quasiparticle picture but could be still amenable to a description in terms of certain one- and two-particle Green functions. 

In fact, had this implicit 
assumption failed as well, it would make any comparison with the 
experimental data (deduced by means of the available one- and two-particle probes) rather problematic. 

To that end, a comparison with the results obtained by other, more traditional, techniques might be helpful for setting up a proper holographic model.
Besides, in order to become a viable practical tool, the holographic
approach would have to be able to reproduce the behavior of 
not just one, but a whole variety of observables, such as 
specific heat, compressibility,
magnetic susceptibility, electrical, thermal, and spin conductivities, etc.
A host of such data on the documented NFL materials is available
and, for the most part, is still awaiting its interpretation.

In the present communication, a number of 
the existing holographic predictions for thermodynamic and kinetic coefficients in the theories dual to the HV geometries
were analysed in the framework of
the scaling theory and with an eye on the general universal 
relations. In the course of such analysis, a number of contradictions between the predictions for, e.g., the conductivity obtained by virtue of the Kubo vs 'membrane paradigm' techniques
were exposed and the related subtleties emphasized.

Providing a solid physical interpretation for
the holographic results is instrumental for ascertaining their true status. 
In the absence of such physical input, the only (obviously, unwanted) 
alternative for the holographic predictions would be to get stuck in the situation where 
any formal result would seem to be (almost) as good as any other one.
Only after having proven to be more than tenuously related to the actual materials, will the holographic approach become a genuine breakthrough in the field of strongly correlated systems. 

The author acknowledges communications 
with C. P. Herzog, W. Witczak-Krempa, J. F. Pedraza, J. Hartong, and A. Mukhopadhyay,  
following the first release of this note, 
as well as hospitality at the Aspen Center for Physics
funded by the NSF under Grant 1066293 and at
the International Institute of Physics in Natal, Brazil 
funded by MCTI and MEC.

\end{document}